# Can we reconstruct a dysarthric voice with the large speech model Parler TTS?

*Ariadna Sanchez, Simon King*

Centre for Speech Technology Research, University of Edinburgh, UK

`ariadna.sanchez@ed.ac.uk, Simon.King@ed.ac.uk`

## Abstract

Speech disorders can make communication hard or even impossible for those who develop them. Personalised Text-to-Speech is an attractive option as a communication aid. We attempt voice reconstruction using a large speech model, with which we generate an approximation of a dysarthric speaker's voice prior to the onset of their condition. In particular, we investigate whether a state-of-the-art large speech model, Parler TTS, can generate intelligible speech while maintaining speaker identity. We curate a dataset and annotate it with relevant speaker and intelligibility information, and use this to fine-tune the model. Our results show that the model can indeed learn to generate from the distribution of this challenging data, but struggles to control intelligibility and to maintain consistent speaker identity. We propose future directions to improve controllability of this class of model, for the voice reconstruction task.

## 1. Introduction

Speech disorders, caused by various neurological disorders, are increasing worldwide. Particularly, over 50 million people have developed a neurodegenerative condition, with the World Health Organisation estimating that it will double in the next 30 years [1]. Eventually, some of these people will become users of a voice output communication aid (VOCA), which uses Text-to-Speech (TTS). Advances in TTS and voice conversion (VC) have enabled the creation of *personalised* VOCAs, producing intelligible speech that maintains some of the speaker's voice characteristics, such as their pitch patterns and accent. Using a VOCA system with a *personalised* voice leads to an improved identity of the self in the user [2]. Several companies currently offer this type of service, including SpeakUnique [3] and VocalID [4].

Recently, TTS has shifted paradigm from training models from scratch to fine-tuning base models, as in other fields like computer vision or natural language processing. These models are now generative, and even though they give good results in terms of naturalness, they still exhibit variability in speaker and accent consistency [5], even for a target speaker with healthy speech. It is not yet known how these models perform for *voice reconstruction*, which we define as synthesising in the voice of a target speaker whose recordings have disordered speech characteristics, but where the synthetic output should "remove" those disordered traits while maintaining speaker qualities, including pitch and accent.

For our experiments, we use the large speech model architecture, Parler TTS [5, 6]. Our goal is to improve intelligibility by focusing on minimising imprecise consonants, a trait associated with dysarthria. Dysarthria is a speech disorder caused by the weakening of the speech production muscles, resulting in slow, slurred speech with articulation difficulties. We create a proof-of-concept model and evaluate whether we can control intelligibility and speaker identity, and thus perform voice reconstruction. Before starting this work, our hypothesis was that, while the model will be able to sample from the distribution of healthy and disordered speech traits represented in the data, it will not be able to control these characteristics and consistently reconstruct a voice.

To the best of our knowledge, this is the first attempt at voice reconstruction with a large speech model. In addition to constructing a proof-of-concept system that is reproducible, we investigate how commonly-used objective metrics, such as Word Error Rate (WER), speaker similarity, and objective mean opinion score (MOS) perform for dysarthric speech. We also propose future work on how to improve the controllability of large speech models for the voice reconstruction task.

## 2. Voice banking and reconstruction

Traditionally, personalised VOCAs have relied on Voice Banking: recording the user's voice prior to disease onset, or in the very early stages, then using these recordings later to build a personalised TTS [7, 8]. This works well, except that, it does not cater for users that already exhibit disorders in their speech.

Voice reconstruction (or "repair") techniques have been proposed for this situation. There are two distinct lines of work in the literature. In one, there are attempts to reconstruct the voice recordings, which can then be used as the data for building a TTS voice. Authors in [9] proposed using signal processing, while others used Voice Conversion to transform healthy speech samples (from other speakers) to the target speaker's characteristics, but without their dysarthric traits [10, 11, 12, 13, 14].

In the other line of work, the reconstruction is attempted within the TTS system. A HMM-based speech synthesis approach [15, 7] involves training an average voice model of similar characteristics (perceived gender and accent) and a user voice model. Repair is achieved by substituting or interpolating parts of the user model with corresponding parts of the average voice model. Initially, this required expert speech therapists to decide which parts to repair, but an automatic process was developed later [16]. More recent work, using deep learning, proposed an augmented reconstruction loss that guides the TTS model to produce healthy speech articulation [17].

## 3. Dataset

### 3.1. Speech Accessibility Project dataset

We used the Speech Accessibility Project (SAP) dataset of English speech [18], designed for improving speaker-independent automatic speech recognition systems, as our source of dysarthric speech. We selected the SAP dataset instead of other widely used datasets because of its higher recording quality, total amount of speakers, and recorded data per speaker [19].

We used the data release from the SAP challenge at Interspeech 2025. We selected Amyotrophic Lateral Sclerosis (ALS) and Cerebral Palsy (CP) etiologies because these conditions are more prone to manifest dysarthric speech. For example, 80-90% of individuals with ALS end up developing severe dysarthria, which requires them to use augmentative and alternative communication systems at some point [20]. We removed annotations in the transcription, keeping only the message of the speaker. We retained both read and spontaneous speech. We took the average of all ratings provided per speaker for 'Imprecise consonants' as a proxy for intelligibility because it is strongly linked to articulation. These ratings were provided by speech therapists on a 1-7 Likert scale, with 1 corresponding to lowest severity, and 7 to highest. Speakers without ratings were discarded. Because the dataset does not provide each speaker's gender, we used an XLSR-53 based automatic gender recognition model[1] [21] to obtain those labels[2].

In total, the curated SAP dataset contains 22.97 hours of speech. It includes 22 speakers with an average of 1 hour of speech (min 0.23 hours, max 2.66 hours, sd 0.47 hours).

### 3.2. Constructing a matched control dataset

In order to fine-tune the model for the voice reconstruction task, we need "healthy" speech along with dysarthric speech, which serves as examples of target articulation. As the SAP dataset does not contain speech of undiagnosed participants, we selected speakers from the English training partition of the Multilingual LibriSpeech (MLS) dataset [22]. The MLS dataset is comprised of audiobook recordings, and is part of the training set for the Parler TTS model we used. Each SAP speaker was matched to a speaker in the MLS dataset by predicted gender and total data length. Given the age population difference between the SAP and MLS datasets, and the potential unreliability of age classifiers, we did not attempt to match speakers by age.

Parler TTS uses text descriptions (prompts) of the speaker and speaking style, so the fine-tuning data needs to be annotated with these: i.e., they are labels on the fine-tuning data. We used the DataSpeech package [5, 6], to do this, for all data above. Our prompts include a description of the intelligibility rating (for the healthy controls[3], 0: 'Extremely good intelligibility'; for the SAP dataset, 1: 'Very good intelligibility', 2: 'Good intelligibility', 3: 'Slightly good intelligibility', 4: 'Slightly poor intelligibility', 5: 'Poor intelligibility', 6: 'Very poor intelligibility', 7: 'Extremely poor intelligibility'). The prompts also include a unique name for each speaker in the dataset according to the predicted gender (e.g. Carla, Jason). The matched control dataset contains a total of 22.94 hours of speech, with the same amount of speakers and data distribution than in the SAP dataset (i.e. 22 speakers, min 0.23 hours, max 2.66 hours, sd 0.47 hours).

---
[1]https://huggingface.co/alefiury/wav2vec2-large-xlsr-53-gender-recognition-librispeech
[2]Gender recognition systems have limitations. They cannot label the gender of a speaker, but only generalise to what their gender might be perceived as. The model we use is trained to produce a binary output, which does not reflect the spectrum of gender identity.
[3]Speakers in the healthy control dataset are likely not to be all equally intelligible, but differentiating intelligibility amongst them is out of scope for this work.

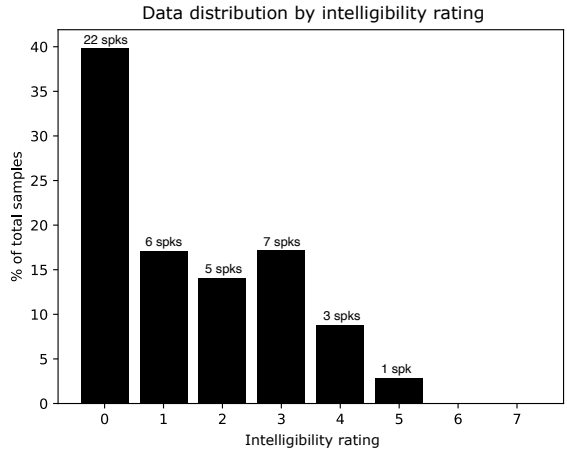

Figure 1: *Distribution of speech samples by intelligibility rating. All control speakers are assumed to have healthy speech and so are given a rating of 0; ratings of 1 - 7 are for speakers in the SAP dataset (dysarthric speech). The number of speakers is indicated above each bar.*

Figure 1 displays the percentage of data per intelligibility rating in the combined dataset, and the total number of speakers for each category. No speakers in the dataset have an articulation rating of 6 or 7. Therefore, the distribution within the SAP portion is skewed towards intelligibility ratings of lower severity (i.e., more intelligible, ratings of 1-3), while the entire control portion has a rating of 0.

## 4. Method

### 4.1. Model

We fine-tuned Parler TTS Mini v0.1 [5, 6] with the combined SAP and matched control data, in a similar fashion to models previously fine-tuned on healthy speech[4]. Sampling rate for all data is 44.1 kHz. Any samples longer than 20 s or shorter than 2 s, or text length over 400 characters, were discarded. We trained for 2 epochs, with the following parameters: batch size of 2, Adam beta1 0.9, Adam beta2 0.99, weight decay 0.01, *constant_with_warmup* learning rate scheduler, which keeps the learning rate high, and warmup steps 50. Using one NVIDIA-A100-SXM4-80GB GPU, fine-tuning took 2 hours.

### 4.2. Inference

We carefully selected 3 speakers (which we refer to here as A, B, and C) from the SAP dataset, based on their intelligibility ratings (1, 2, and 5) and on their distinct speaker traits, which should facilitate assessing if speaker identity is consistent. We also selected 3 speakers from the matched control data (which we refer to here as D, E, and F). For each of the 6 speakers, we synthesised test sentences at each intelligibility rating, from that speaker's true intelligibility rating, in steps of 1, up to a rating of 0 (i.e., healthy). For speakers D, E, and F, this means

---
[4]https://github.com/huggingface/parler-tts/tree/main/training#3-training

only synthesising at a rating of 0. That is, we do not attempt to *degrade* any speaker's intelligibility.[5]

As test material, we selected 2500 sentences from the Clarity 2021 challenge devset[6] [23]. We chose the Clarity devset because the sentences contain 7-10 words, were checked manually for acceptable grammar and vocabulary [23], and are not in either the original training data or the fine-tuning data for Parler.

For each speaker, we synthesised speech samples using various prompts, ranging from the speaker's original intelligibility rating (e.g., 'poor intelligibility') to fully-healthy speech (i.e., 'extremely good intelligibility'). The prompt template was: *"[Name of the speaker] speaks with a monotone and medium-pitched voice, with [intelligibility rating], and is delivering [his/her] speech at a normal speed in a confined environment."*. For each speaker and intelligibility rating, we only generated a single rendition of each sentence. Note that Parler TTS is a generative model, so each rendition will randomly vary. How best to sample from such a model is an active line of research [24].

### 4.3. Evaluation

We evaluated the model with a combination of objective measures and expert listening[7]. Subjective evaluation by non-expert listeners was not used because such listeners will not be familiar with the voice reconstruction task, and will generally have no experience in listening to dysarthric speech. As a point of comparison, previous work has shown that listeners struggle to assess how similar a speaker is when speaking in different languages [25]. Using subjective evaluations to assess voice reconstruction poses very interesting questions, but these are out of scope for now.

We evaluated intelligibility, speaker similarity, and naturalness using objective measures. We employed widely-used state-of-the-art tools - Whisper large [26] to measure intelligibility in terms of WER, Resemblyzer [27] for speaker similarity, and UTMOS [28] for objective MOS naturalness.

But before deploying these tools on the synthetic speech, we first investigated how they perform on natural dysarthric speech recordings, and on the healthy control speech data. We expected that WER will increase as intelligibility decreases. Ideally, speaker similarity and UTMOS should not be influenced by the intelligibility rating of the speaker, since their speech is – by definition – perfectly similar and entirely natural. However, UTMOS is trained to replicate listeners' ratings of synthetic speech "naturalness", which is not the same thing as human-like. Average WER was computed over all the data for each speaker. We obtained a speaker embedding for each speaker from 50 speech samples and compared that against all other samples from that speaker. We obtained a MOS naturalness score per speech sample in the fine-tuning dataset.

For the synthesised samples, we performed a similar analysis to the one described above for natural speech. The only difference was for speaker similarity, where we obtained a speaker embedding from 50 samples of recordings of natural speech, then calculated similarity scores for all synthetic samples.

---
[5]Although this might have useful applications such as augmenting data for training an ASR or speaker verification model, for example.
[6]clarity_CEC2_core.v1_0.tgz from https://claritychallenge.org/clarity/recipes/cec2/README.html
[7]In our case, we define expert listening as the subjective evaluation conducted by the authors.

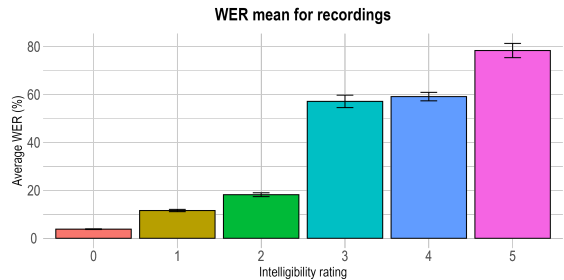

Figure 2: *Average WER (%) for the data used to fine-tune Parler TTS, broken down by intelligibility rating. Note that each intelligibility rating band has a different number of speakers (see Figure 1).*

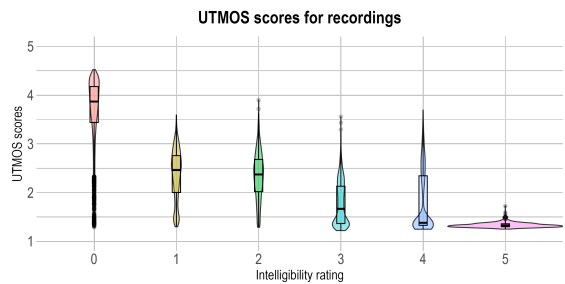

Figure 3: *UTMOS naturalness scores for the fine-tuning dataset, broken down by intelligibility rating. The scores were calculated using the official implementation [28]. Note that each intelligibility rating band has a different number of speakers (see Figure 1).*

## 5. Results

### 5.1. Performance of objective measures on recorded speech

Figure 2 shows the average WER for all speakers in the training set, broken down by intelligibility rating. WER is very low for the healthy control speakers, but considerably higher for all speakers in the SAP data. It is, of course, likely that MLS English (the source of our healthy control speakers) is in the Whisper training data. Nevertheless, the results for the SAP speakers are consistent with the fact that ASR systems perform poorly for dysarthric speech [29]. This means that the intelligibility rating included in the Parler prompts for the fine-tuning data does indeed correlate with speaker intelligibility, and the model should therefore respond to this during inference.

Figure 3 shows the distribution of UTMOS scores for the fine-tuning data, again broken down by intelligibility rating. UTMOS is evidently highly sensitive to the speech type, giving low scores to the unseen type: dysarthric speech. Even for relatively intelligible speakers in the SAP dataset, scores are generally below 3. UTMOS does not generalise well to this type of speech. Given this, we cannot use UTMOS as an objective measure for our synthetic speech. We will rely on expert listening instead.

Finally, Figure 4 shows the distribution of speaker similarity (remembering that these are comparisons of recorded speech with other examples of recorded speech from the same speaker). Resemblyzer embeddings are robust to this unseen type of speech (dysarthric). The range of scores for the healthy controls and for each band of speakers across the different intelligibility ratings, are quite similar. This indicates that this measure should be appropriate for use on our synthetic speech.

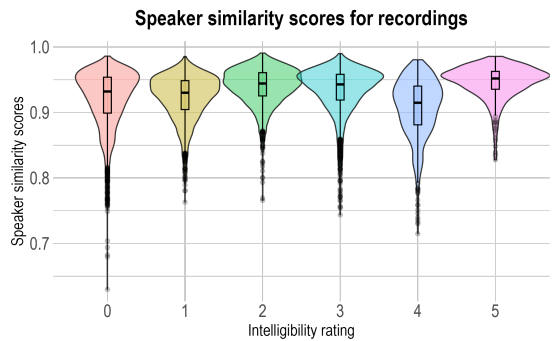

Figure 4: *Speaker similarity scores for the fine dataset, broken down by intelligibility rating. Scores were calculated following the official implementation of Resemblyzer [27]. Note that each intelligibility rating band contains a different number of speakers (Figure 1).*

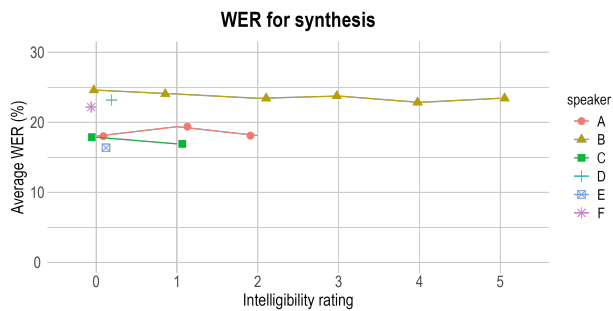

Figure 5: *Average WER (%) for synthesised speech at different levels of intelligibility (which is specified as part of the Parler prompt during inference), for each of the 6 test speakers. (Points are slightly jittered for better visualisation.)*

### 5.2. Results from objective measures for synthetic speech

We calculated average WER and speaker similarity scores for dysarthric speakers (A, B, and C), and healthy control speakers (D, E, and F).

Figures 5 and 6 show the average WER and speaker similarity for each speaker at different intelligibility ratings. In the case of WER, we observe very little variation from ratings 5 to 0, with WER generally being around 20-25% for every speaker, including those from the matched control dataset. Similarly, speaker similarity scores do not vary much across different intelligibility ratings.

### 5.3. Results from expert listening for synthetic speech

Expert listening revealed that the model is *sampling* from the various levels of intelligibility and also from speaker identity,

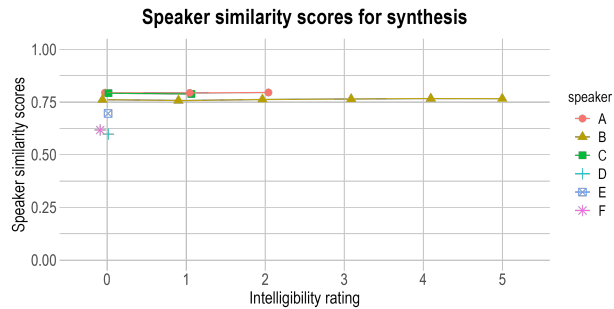

Figure 6: *Same as Figure 5 except for average speaker similarity scores.*

even for a fixed prompt that specifies the desired values[8]. This output variability explains why Figures 5 and 6 show similar average WER and speaker similarity across speakers and intelligibility ratings.

Parler, like most state-of-the-art TTS models is generative, so we must expect some variation in the output. It is therefore worth listening to a substantial sample of model output. When doing so, we do in fact find instances where the synthesised sample sounds similar to the target speaker *and* has higher intelligibility than that speaker's recorded speech (although the model never *fully* reconstructs this all the way to what we would expect in a healthy voice).

## 6. Discussion and conclusion

Our interpretation of the results is that the model is *capable* of voice reconstruction. Currently, the degree of control offered by the natural language prompt is too weak to provide consistent results during inference. This is probably because of the way Parler uses this prompt: the entire prompt is embedded by a pre-trained NLP model, then attended to by the speech language model. A more direct form of control, especially for a categorical label such as speaker, is likely to work better.

The Parler checkpoint that we used was trained not with speaker identities but just general speaker descriptions like "A woman is speaking...", in the prompt. It is likely that this is a limiting factor when fine-tuning this model to generate *fixed* speaker identities. Another possibility would be to include more specific information about the speaker's dysarthria, beyond the *imprecise consonant* assessment that we used in this proof of concept. SAP includes ratings for harshness, speech monotony, and so on.

Another limiting factor, common to all work with dysarthric speech, is the limited amount of data available per speaker, and the limited number of speakers available. One option here might be data augmentation: modify healthy speech in the articulatory domain to produce speech samples with dysarthric traits, by utilising acoustic-to-articulatory inversion models [30, 31]. A clear finding of our work is the poor reliability of some tools used as objective measures, when applied to dysarthric speech.

Finally, there is not yet a standard protocol for evaluating *voice reconstruction*. Proper evaluation is only possible in co-operation with VOCA users, their families and friends, and speech therapists.

---

[8]https://ariadnasc.github.io/parlertts_voice_reconstruction

## 7. Acknowledgements

This work was supported by the UKRI Centre for Doctoral Training in Natural Language Processing, funded by UKRI (grant EP/S022481/1).

We want to thank Cassia Valentini-Botinhao, Cameron Scott and Artemis Deligianni for their insightful comments and feedback on this work. We also want to thank Mark Hasegawa-Jonhson and collaborators for creating and sharing the Speech Accessibility Project dataset with the research community, and to the participants for donating their voice.

## 8. References

[1] Cleveland Clinic, Dec 2024. [Online]. Available: https://my.clevelandclinic.org/health/diseases/24976-neurodegenerative-diseases

[2] J. Murphy, ""I prefer contact this close": Perceptions of AAC by people with motor neurone disease and their communication partners," *Augmentative and Alternative Communication*, vol. 20, no. 4, pp. 259–271, 2004. [Online]. Available: https://doi.org/10.1080/07434610400005663

[3] SpeakUnique. [Online]. Available: https://www.speakunique.co.uk/

[4] C. Jreige, R. Patel, and H. T. Bunnell, "VocaliD: Personalizing text-to-speech synthesis for individuals with severe speech impairment," in *Proceedings of the 11th international ACM SIGACCESS conference on Computers and accessibility*, 2009, pp. 259–260.

[5] Y. Lacombe, V. Srivastav, and S. Gandhi, "Parler-TTS," https://github.com/huggingface/parler-tts, 2024.

[6] D. Lyth and S. King, "Natural language guidance of high-fidelity text-to-speech with synthetic annotations," 2024. [Online]. Available: https://arxiv.org/abs/2402.01912

[7] J. Yamagishi, C. Veaux, S. King *et al.*, "Speech synthesis technologies for individuals with vocal disabilities: Voice banking and reconstruction," *Acoustical Science and Technology*, vol. 33, no. 1, pp. 1–5, 2012.

[8] S. Judge and N. Hayton, "Voice banking for individuals living with MND: A service review," *Technology and Disability*, vol. 34, no. 2, pp. 113–122, 2022.

[9] F. Rudzicz, "Production knowledge in the recognition of dysarthric speech," Ph.D. dissertation, University of Toronto, 2011.

[10] T. Mills, H. T. Bunnell, and R. Patel, "Towards personalized speech synthesis for augmentative and alternative communication," *Augmentative and Alternative Communication*, vol. 30, no. 3, pp. 226–236, 2014.

[11] S. Yang and M. Chung, "Improving dysarthric speech intelligibility using cycle-consistent adversarial training," in *Proceedings of the 13th International Joint Conference on Biomedical Engineering Systems and Technologies*. SCITEPRESS-Science and Technology Publications, 2020.

[12] D. Wang, J. Yu, X. Wu *et al.*, "End-to-end voice conversion via cross-modal knowledge distillation for dysarthric speech reconstruction," in *ICASSP 2020-2020 IEEE International Conference on Acoustics, Speech and Signal Processing (ICASSP)*. IEEE, 2020, pp. 7744–7748.

[13] M. Purohit, M. Patel, H. Malaviya *et al.*, "Intelligibility improvement of dysarthric speech using MMSE discogan," in *2020 International Conference on Signal Processing and Communications (SPCOM)*. IEEE, 2020, pp. 1–5.

[14] Y. Zhao, M. Song, Y. Yue *et al.*, "Personalizing TTS voices for progressive dysarthria," in *2021 IEEE EMBS International Conference on Biomedical and Health Informatics (BHI)*. IEEE, 2021, pp. 1–4.

[15] C. Veaux, J. Yamagishi, and S. King, "Using HMM-based speech synthesis to reconstruct the voice of individuals with degenerative speech disorders." in *Interspeech*, 2012, pp. 967–970.

[16] ——, "A comparison of manual and automatic voice repair for individual with vocal disabilities," in *Proceedings of SLPAT 2015: 6th Workshop on Speech and Language Processing for Assistive Technologies*, 2015, pp. 130–133.

[17] Y. Tian, J. Li, and T. Lee, "Creating personalized synthetic voices from articulation impaired speech using augmented reconstruction loss," in *ICASSP 2024-2024 IEEE International Conference on Acoustics, Speech and Signal Processing (ICASSP)*. IEEE, 2024, pp. 11 501–11 505.

[18] M. Hasegawa-Johnson, X. Zheng, H. Kim *et al.*, "Community-supported shared infrastructure in support of speech accessibility," *Journal of Speech, Language, and Hearing Research*, vol. 67, no. 11, pp. 4162–4175, 2024.

[19] G. Schu, P. Janbakhshi, and I. Kodrasi, "On using the UA-speech and TORGO databases to validate automatic dysarthric speech classification approaches," in *ICASSP 2023-2023 IEEE International Conference on Acoustics, Speech and Signal Processing (ICASSP)*. IEEE, 2023, pp. 1–5.

[20] D. R. Beukelman, J. Childes, T. Carrell *et al.*, "Perceived attention allocation of listeners who transcribe the speech of speakers with amyotrophic lateral sclerosis," *Speech Communication*, vol. 53, no. 6, pp. 801–806, 2011.

[21] A. Conneau, A. Baevski, R. Collobert *et al.*, "Unsupervised cross-lingual representation learning for speech recognition," in *Interspeech 2021*, 2021, pp. 2426–2430.

[22] V. Pratap, Q. Xu, A. Sriram *et al.*, "MLS: A large-scale multilingual dataset for speech research," in *Interspeech 2020*, 2020, pp. 2757–2761.

[23] S. Graetzer, J. Barker, T. J. Cox *et al.*, "Clarity-2021 challenges: Machine learning challenges for advancing hearing aid processing," in *Proceedings of the Annual Conference of the International Speech Communication Association, INTERSPEECH*, vol. 2, 2021.

[24] A. Adigwe, S. Wallbridge, Z. Tu *et al.*, "Can we cherry pick? investigating multiple rendition from a generative speech model," in *ICASSP*, 2025.

[25] M. Wester and R. Karhila, "Speaker similarity evaluation of foreign-accented speech synthesis using hmm-based speaker adaptation," in *2011 IEEE International Conference on Acoustics, Speech and Signal Processing (ICASSP)*. IEEE, 2011, pp. 5372–5375.

[26] A. Radford, J. W. Kim, T. Xu *et al.*, "Robust speech recognition via large-scale weak supervision," 2022. [Online]. Available: https://arxiv.org/abs/2212.04356

[27] Resemble-Ai, "Resemble-ai/resemblyzer: A python package to analyze and compare voices with deep learning," 2020. [Online]. Available: https://github.com/resemble-ai/Resemblyzer

[28] T. Saeki, D. Xin, W. Nakata *et al.*, "UTMOS: UTokyo-SaruLab system for VoiceMOS Challenge 2022," in *Interspeech 2022*, 2022, pp. 4521–4525.

[29] H. Christensen, S. P. Cunningham, C. Fox *et al.*, "A comparative study of adaptive, automatic recognition of disordered speech." in *Interspeech*. Portland, 2012, pp. 1776–1779.

[30] P. Wu, L.-W. Chen, C. J. Cho *et al.*, "Speaker-independent acoustic-to-articulatory speech inversion," in *ICASSP*, 2023.

[31] Y. Liu, B. Yu, D. Lin *et al.*, "Fast, high-quality and parameter-efficient articulatory synthesis using differentiable DSP," in *2024 IEEE Spoken Language Technology Workshop (SLT)*. IEEE, 2024, pp. 711–718.